\documentclass[fleqn,10pt]{wlscirep}
\usepackage[utf8]{inputenc}
\usepackage[T1]{fontenc}
\usepackage{lineno}
\usepackage{multirow}   
\usepackage{tabularx} 
\usepackage{cite}
\usepackage{hyperref}
\usepackage{colortbl} 
\usepackage{xcolor}   

\title{Association in Facial Phenotype, Gene, Disease: A Dataset for Explainable Rare Genetic Diseases Diagnosis}

\author[1,$\dag$]{Jie Song}
\author[1,$\dag$]{Mengqiao He}
\author[1,*]{Shumin Ren}
\author[1,*]{Bairong Shen}
\affil[1]{Department of Ophthalmology and Institutes for Systems Genetics, Frontiers Science Center for Disease-related Molecular Network, West China Hospital, Sichuan University, Chengdu 610212, China}

\affil[*]{corresponding authors: Bairong Shen (Bairong.shen@scu.edu.cn), Shumin Ren (renshumin@wchscu.cn)}

\affil[$\dag$]{these authors contributed equally to this work}

\begin{abstract}
Many rare genetic diseases exhibit recognizable facial phenotypes, which are often used as diagnostic clues. 
However, current facial phenotype diagnostic models, which are trained on image datasets, have high accuracy but often suffer from an inability to explain their predictions, which reduces physicians' confidence in the model output.
In this paper, we constructed a dataset, called FGDD, which was collected from 509 publications and contains 1147 data records, in which each data record represents a patient group and contains patient information, variation information, and facial phenotype information.
To verify the availability of the dataset, we evaluated the performance of commonly used classification algorithms on the dataset and analyzed the explainability from global and local perspectives. 
FGDD aims to support the training of disease diagnostic models, provide explainable results, and increase physicians' confidence with solid evidence. 
It also allows us to explore the complex relationship between genes, diseases, and facial phenotypes, to gain a deeper understanding of the pathogenesis and clinical manifestations of rare genetic diseases. 

\end{abstract}
\begin{document}

\flushbottom
\maketitle

\thispagestyle{empty}

\section*{Background \& Summary}

More than 6$\%$ of the world's population is affected by rare genetic diseases\cite{ferreira2019burden}, and the latest Orphanet\cite{orphanet} and OMIM\cite{OMIM} databases show that there are currently at least 7000 rare genetic diseases.
Among these, many rare genetic diseases have recognizable facial phenotypic features, and facial phenotypes are often used as a basis for diagnosis\cite{Crouzon, Williams, CSS,craniofacial}. 
In recent studies, the diagnosis of rare genetic diseases through computer vision techniques has reached the level of clinical experts\cite{thevenot2017survey, DeepGestalt, bannister2022deep, hallgrimsson2020automated, jin2020deep, pooch2020computational, singh2018detection}.

For example, DeepGestalt\cite{DeepGestalt}, trained on a private dataset of more than 17,000 images, can identify the correct disease in 502 different images, achieving a top 10 accuracy of 91$\%$. 
But like many AI models, DeepGestalt was unable to explicitly explain its predictions or provide information about which facial features drove the diagnosis. 

AI models rely heavily on data, and the GestaltMatcher Database (GMDB)\cite{GMDB} is the only publicly available dataset in the field, containing 10,189 frontal images of 7,695 patients with 683 diseases. 
However, the models trained on GMDB suffer from the same inability to explain their predictions as DeepGestalt, but in the field of medicine, explainability is crucial\cite{yoon2022machine, vellido2020importance, teng2022survey}, so an explainable dataset in this field is urgently needed. 

Compared to the GMDB image format, tabular format is naturally more explainable.
Tabular dataset with clear meanings and units, the relationship between data is more direct and easier to be understood by humans\cite{rudin2022interpretable}. For instance, when we see "age" and "disease" in the table, we can intuitively understand the correlation between them.
Moreover, tabular dataset consists of numerical values and categories that can be directly understood by humans, while GMDB image data consists of pixel values that do not have direct semantic information.
Furthermore, some logical rules can be found in tabular data, such as "c.6726\_6730del; p.Leu2243Serfs*8 in Exon 20 cause Coffin-Siris syndrome 1", whereas for image data, what the model learns tends to be high-level feature combinations.

Therefore, We propose a new tabular dataset, FGDD, which contains 1147 data records, 197 associated genes, 437 associated phenotypes, and 211 associated diseases, of which 689 data records have disease labels.
FGDD was constructed by retrieving publications from Human Phenotype Ontology (HPO\cite{HPO})-generated terms, and then identifying facial phenotypes-gene-disease associations from these publications.

The FGDD is primarily used for facial phenotype analysis of rare genetic diseases. 
It serves multiple purposes, including training explainable diagnostic models, conducting in-depth analysis of the complex relationships between genes, diseases, and facial phenotypes, and uncovering additional potential associations and patterns.

Our contributions can be summarized as follows:
\begin{itemize}
    \item{We propose a new dataset, FGDD, for facial phenotype analysis of rare genetic diseases, which can be used not only for training explainable diagnostic models but also for in-depth analysis of the complex gene-disease-facial phenotype relationships and for mining more potential associations and patterns. }
    \item{We have conducted extensive benchmarking on our dataset, and commonly used machine learning models can achieve up to 81$\%$ accuracy and provide clinical support. }
    \item{We conducted an explainability analysis from both local and global perspectives, and the models trained on FGDD can provide explainable predictions that can be corroborated by relevant research, and enhance physicians' confidence in the model results, which is particularly important in the medical field. }
\end{itemize}

\begin{figure}[htbp]
    \centering
    \includegraphics[width=6.8in]{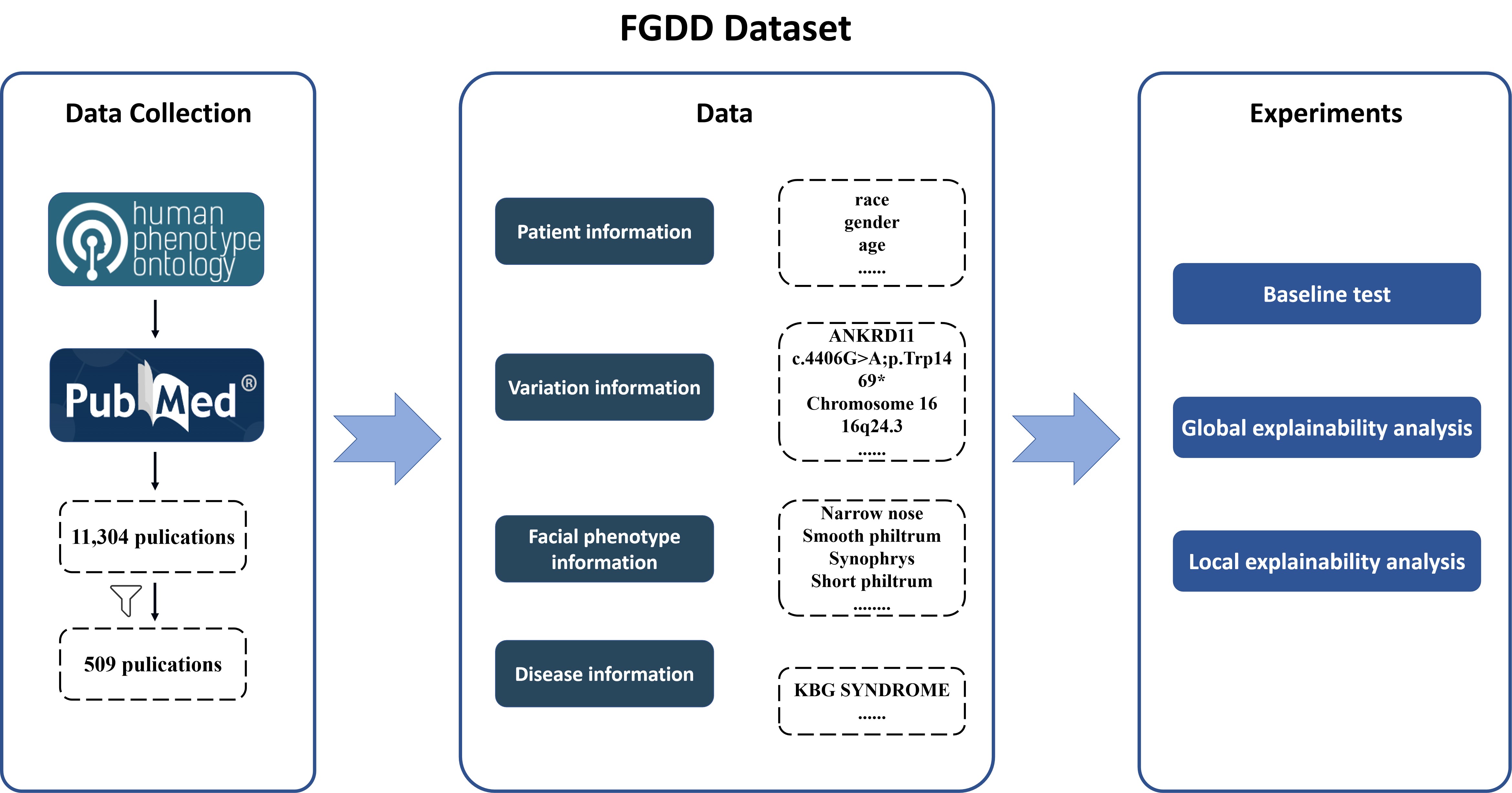}
    \caption{\textbf{The overall process of this study. }}
    \label{fig graph abstract}
\end{figure}

An overview of this study is shown in Fig.\ref{fig graph abstract}. FGDD dataset is available at \href{https://figshare.com/s/89093de15415a773b4ba}{https://figshare.com/s/89093de15415a773b4ba}. 
All codes are publicly available at \href{https://github.com/zhelishisongjie/FGDD}{https://github.com/zhelishisongjie/FGDD} and can be audited, copied, and reused.

\section*{Methods}
\subsection*{Data collection}
We utilized facial phenotype concepts and synonyms from the HPO\cite{HPO} as primary keywords, which were then combined with terms related to genetics and chromosomal variation using the Entrez programming utilities tool\cite{esearch} to construct our search queries.
11,304 publications are retrieved, after screening, 509 publications are involved in this study.

We collected data from the publications using a combination of automated entity recognition tools\cite{phenotagger, pubtator} and manual labor. 
Disease ID was standardized by the \href{https://www.omim.org/}{OMIM}\cite{OMIM} database, Facial phenotypes were standardized by \href{https://hpo.jax.org/app/}{HPO}\cite{HPO}, and gene ID was standardized by the \href{https://www.ncbi.nlm.nih.gov/gene}{NCBI Gene}\cite{NCBI} database. 
In the end, data integration and data rechecking were performed. The process of data collection is shown in Fig. \ref{fig process}, \textbf{the screening details are provided in Fig. S1 in the supplementary materials}.

\subsection*{Implementation details}
For all experiments, we use a split of 70$\%$ and 30$\%$ for the train and test sets, respectively. 
All algorithms used the default settings, so there is no separate validation set to adjust the parameters.
All experiments can be simply reproduced using our codes.

\section*{Data Records}
We propose a new dataset, FGDD, for facial phenotype analysis of rare genetic diseases, which can be used not only for interpretable clinical diagnostic support but also for in-depth analysis of the complex gene-disease-facial phenotype relationships and for mining more potential associations and patterns.

FGDD contains 1147 data records, 197 associated genes, 437 associated phenotypes, and 211 associated diseases, of which 689 data records have disease labels.
The data resources described in this paper are freely and openly available at \href{https://figshare.com/s/89093de15415a773b4ba}{https://figshare.com/s/89093de15415a773b4ba}.
Table \ref{tab overview} provides an overview of the files and datasets stored in figshare. 
All python codes can be audited, replicated, and reused to produce alternative analyses.

\section*{Technical Validation}

\subsection*{Facial phenotype-Gene-Disease relationship analysis}
\begin{figure}[htbp]
    \centering
    \includegraphics[width=6.4in]{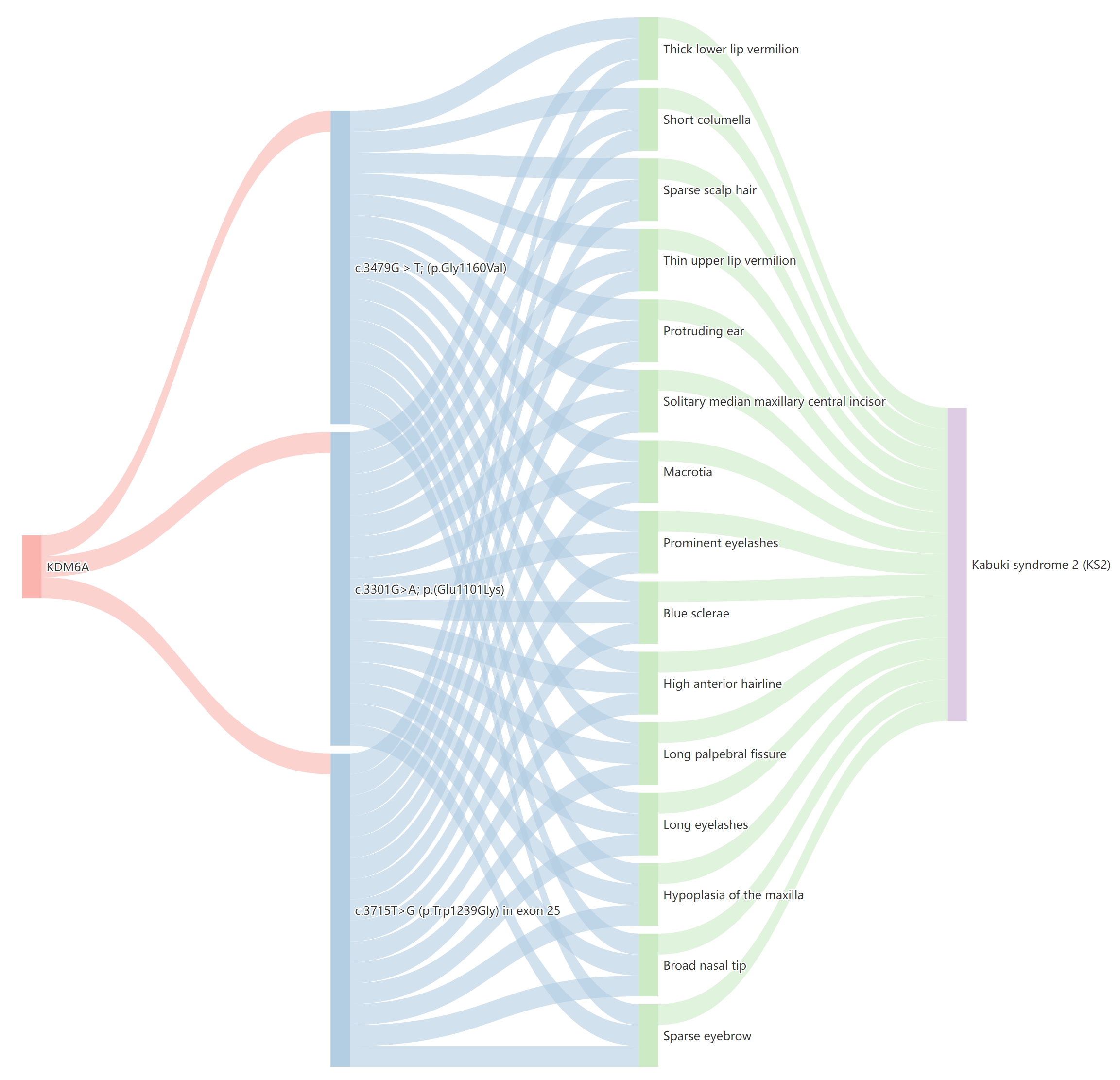}
    \caption{\textbf{Sankey diagram analysis of KDM6A. }The KDM6A gene in the first layer, variant details in the second layer, facial phenotypes in the third layer, and rare genetic diseases in the fourth layer. }
    \label{fig sankey}
\end{figure}

The Sankey diagram shown in Fig. \ref{fig sankey} illustrates the manner in which genetic variations influence facial phenotype, as well as the complex relationship between these variations and underlying diseases. This helps us to comprehend the pathogenesis of genetic diseases and provides valuable reference information for clinical diagnosis and treatment.

\subsection*{Baselines}
We tested the performance of common classification algorithms by splitting the training and test sets using a ratio of 7:3. 
All algorithms used the default settings. The results are shown in Table \ref{tab baseline}.

\subsection*{Explainability analysis}
\textbf{Global explanation}\\
Global explanation is concerned with understanding the logic of the whole model and tries to explain how the model is obtained through learning. Here we focus on feature importance analysis from both coarse-grained and fine-grained perspectives.
Coarse-grained features divide features into three broad categories: patients, variations, and phenotypes, and analyze the importance of all three. Fine-grained features analyze the three categories individually.

\begin{figure}[htbp]
    \centering
    \includegraphics[width=5.0in]{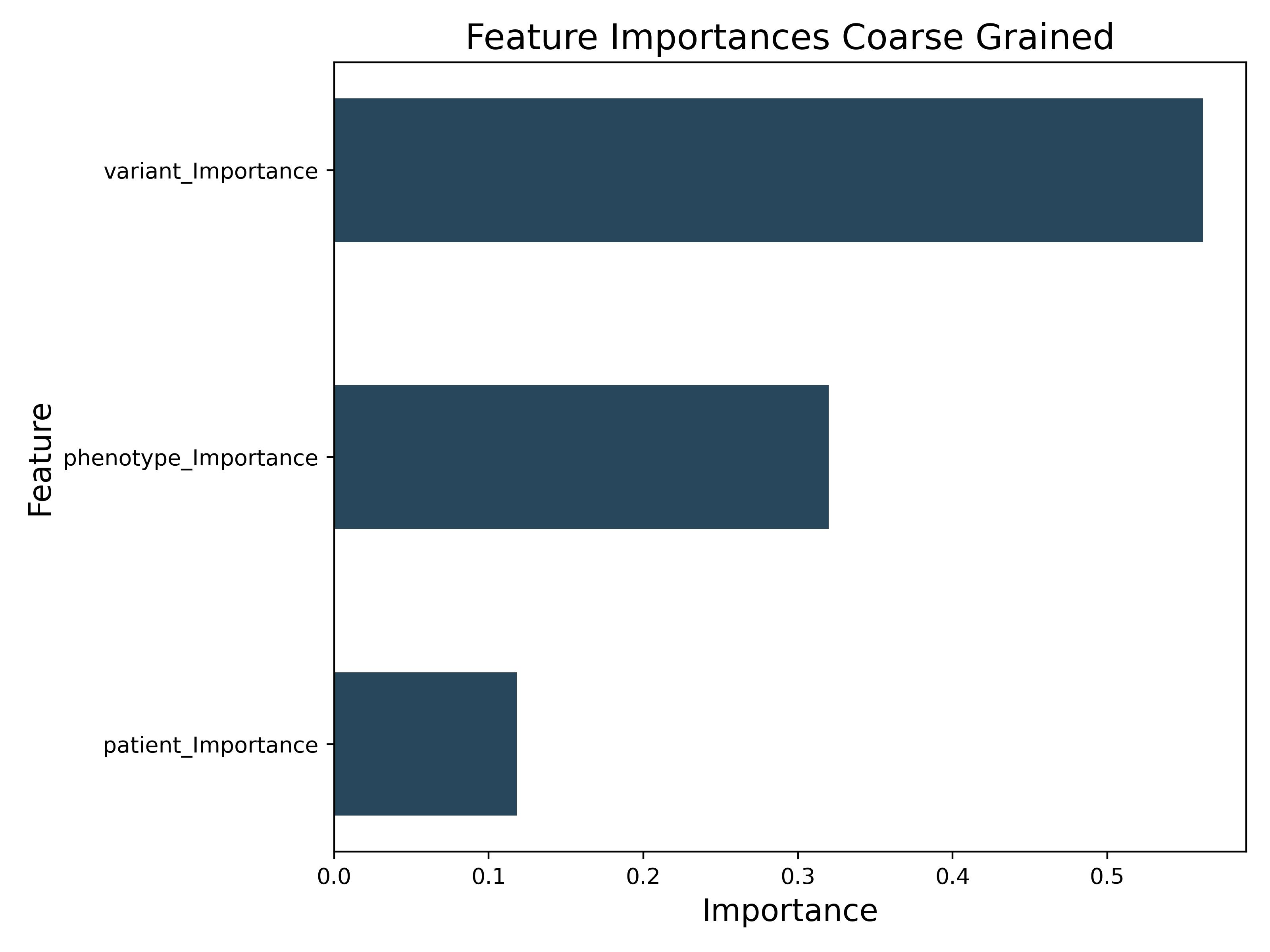}
    \caption{\textbf{coarse-grained features importance. }}
    \label{fig coarse-grained}
\end{figure}

In Fig. \ref{fig coarse-grained} patient information plays a minor role in this diagnostic process. 
This may imply that the pathogenesis of the disease is due to genetic factors more than to individual factors such as environment or lifestyle habits. The variation and facial phenotype are the key diagnostic factors, and genomic analysis should be emphasized in the diagnostic process along with facial phenotypic features to make a more accurate diagnosis. 
Further studies can explore the genotype-phenotype correlation mechanism to better understand the pathogenesis of this rare genetic disease.

\begin{figure*}[htbp]
    \centering
    \includegraphics[width=6.9in]{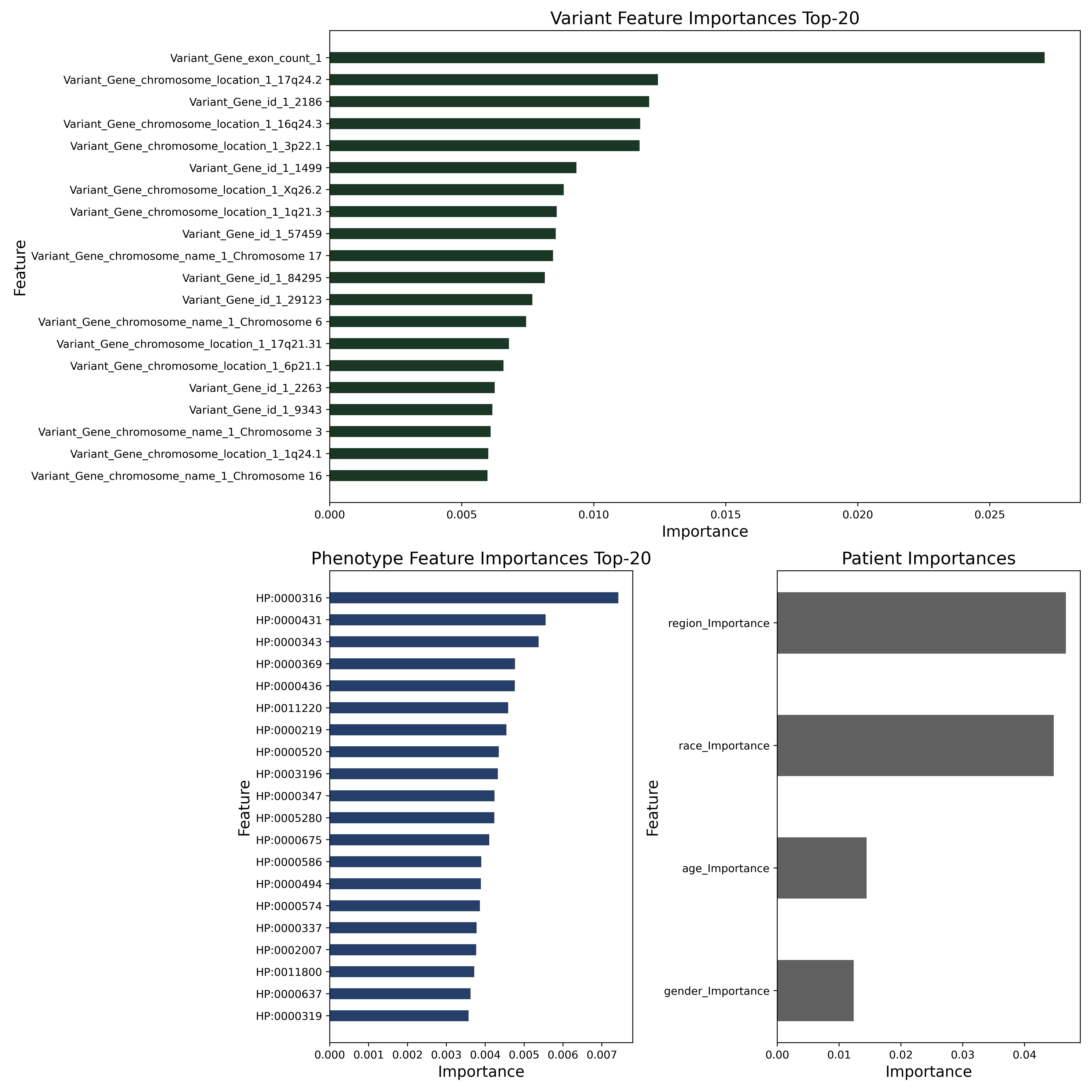}
    \caption{\textbf{Fine-grained features importance. }}
    \label{fig fine-grained}
\end{figure*}

In Fig. \ref{fig fine-grained}, variant gene features play a key role in disease diagnosis. 
Features such as exon count and chromosomal location are essential for accurate disease identification. 
This indicates that genomic analysis is an important basis for diagnosis. Facial phenotype features also contribute to disease diagnosis. 
Several facial phenotype-related clinical features provide valuable information for differentiating between different diseases. 
Physicians need to evaluate the patient's facial phenotypes thoroughly. Background information about the patients themselves, such as geographic region, ethnicity, etc., also has an impact on the diagnosis of certain diseases. 
This implies that some diseases are often related to population or geographic factors and that individual differences in patients need to be taken into account.

\textbf{Local explanation}\\
Local explanations are concerned with explaining the reasons for individual predictions or decisions, answering the question “Why did the model make this prediction for this sample?” Here, we explain individual patient predictions based on SHAP\cite{shap}, a cooperative game-theoretic feature importance calculation method.

\begin{figure*}[htbp]
    \centering
    \includegraphics[width=6.8in]{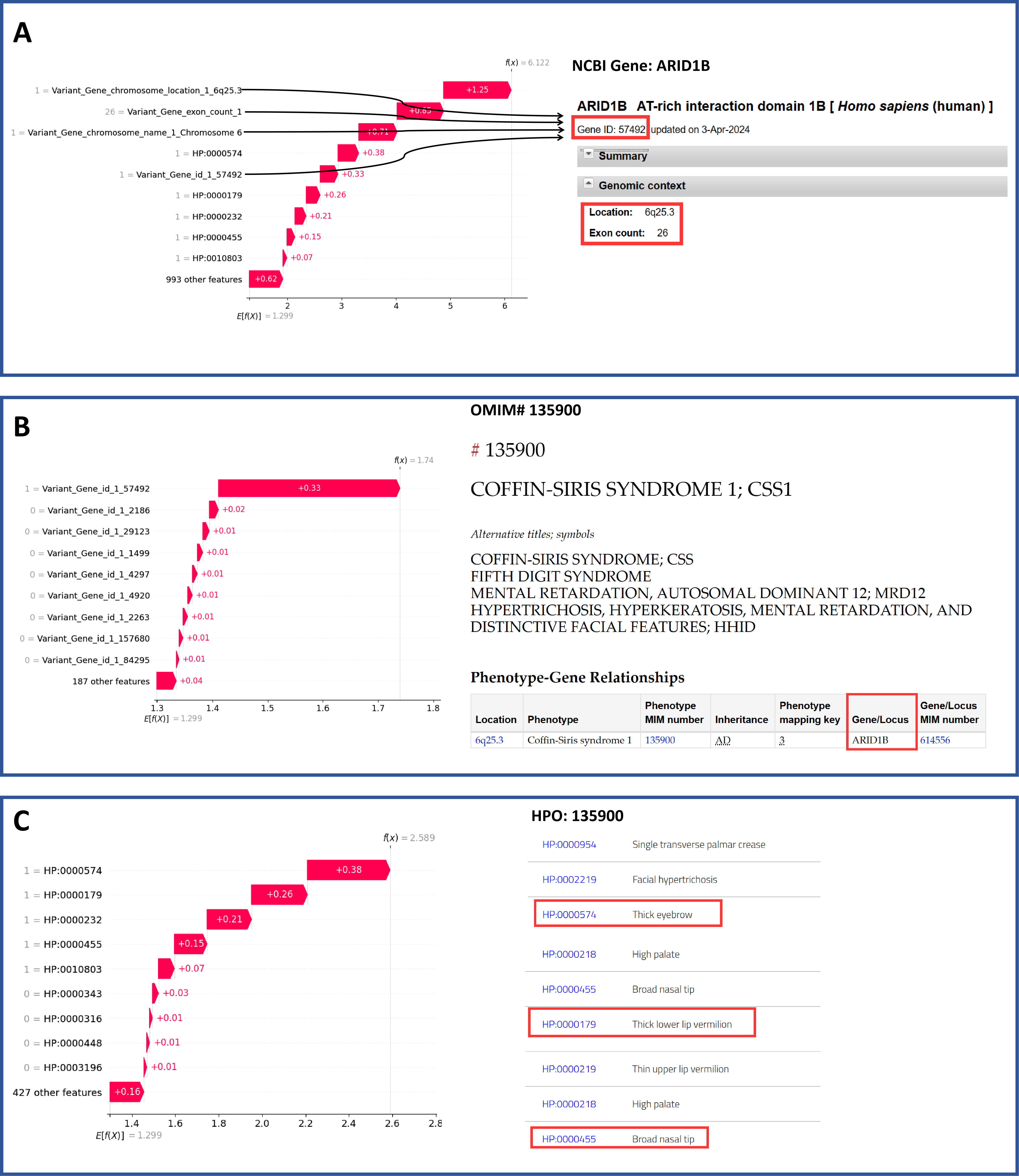}
    \caption{\textbf{Explainability analysis of model single prediction. }}
    \label{fig local}
\end{figure*}

Fig. \ref{fig local} shows a local explanation of a patient with a model diagnosis of COFFIN-SIRIS SYNDROME 1 (OMIM$\#$ 135900). 

Fig. \ref{fig local}A shows the effect of all features on individual prediction results using SHAP analysis, among which the gene ARID1B(57492) has the greatest effect on the prediction results of the model, searching for ARID1B(\href{https://www.ncbi.nlm.nih.gov/gene/57492}{URL}) in the NCBI Gene database, we can find that chromosome and exon count are all ARID1B-related information in the Figure. 

Fig. \ref{fig local}B shows the effect of genes on individual predictions in the SHAP analysis, with ARID1B having the largest effect, leading to speculation about the basis of the model's predictions - that ARID1B plays a critical role in the pathogenesis of COFFIN-SIRIS SYNDROME 1. In order to verify the plausibility of this speculation, search for COFFIN-SIRIS SYNDROME 1(\href{https://www.omim.org/entry/135900}{URL}) on OMIM and find that it is indeed associated with ARID1B. This is also supported by the relevant research\cite{santen2012mutations, van2019arid1b, vasko2021genotype}.

Fig. \ref{fig local}C shows the effect of facial phenotypes on individual predictions in the SHAP analysis, HP:0000574, HP:0000179, HP:0000232, HP:0000455, HP:0010803 have the greatest impact on the predictions, where HP:0000574, HP:0000179, HP:0000455 in the HPO(\href{https://hpo.jax.org/app/browse/disease/OMIM:135900}{URL}) can all be found associated with COFFIN-SIRIS SYNDROME 1. 
Although HPO does not directly mention HP:0000232, and HP:0010803 in relation to COFFIN-SIRIS SYNDROME 1, this does not mean that there is no connection between them. 

HP:0000232, and HP:0010803 for "Everted lower lip vermilion" and "Everted upper lip vermilion", respectively. HPO does not mention their correlation with COFFIN-SIRIS SYNDROME 1. But the HPO explicitly mentions that "Thick lower lip vermilion" and "Thin upper lip vermilion" are related to COFFIN-SIRIS SYNDROME 1 and ARID1B,  and these four facial phenotypes are so similar that they are even categorized in the HPO in the same tier: “Abnormality of upper lip vermillion” and “Abnormality of upper lip vermillion”, perhaps they share common underlying mechanisms and overlapping clinical features, indirectly related to COFFIN-SIRIS SYNDROME 1 and ARID1B, which were not revealed before and may provide new clues and directions for future research and diagnosis. 

Based on the above observations, the model predictions are based on relevant evidence and may find some implicit and unexposed relationships that can help us to deeply understand the pathogenesis and clinical manifestations of rare genetic diseases. 
\textbf{More explanatory analyses of the predictions are provided in Fig. S8-S9 in the supplementary appendix. } \\

\section*{Usage Notes}
The required packages, data pre-processing, specific parameters, codes, and validation methods are all described in detail at
\href{https://github.com/zhelishisongjie/FGDD}{https://github.com/zhelishisongjie/FGDD}. 

\section*{Code availability}

All codes were publicly available at \href{https://github.com/zhelishisongjie/FGDD}{https://github.com/zhelishisongjie/FGDD}.

\bibliography{sample}

\section*{Acknowledgements}
The research was supported by National Supercomputing Center in Chengdu.

\section*{Author contributions statement}

Jie Song conducted data collection, validation, result analysis, and manuscript writing. 
Mengqiao He conducted most of the data collection. 
Shumin Ren revised the manuscript. 
Bairong Shen guided the overall process and revised the manuscript.
All authors reviewed the manuscript.

\section*{Competing interests}

The authors declare no competing interests.

\section*{Figures \& Tables}


\begin{figure}[htbp]
    \centering
    \includegraphics[width=4.8in]{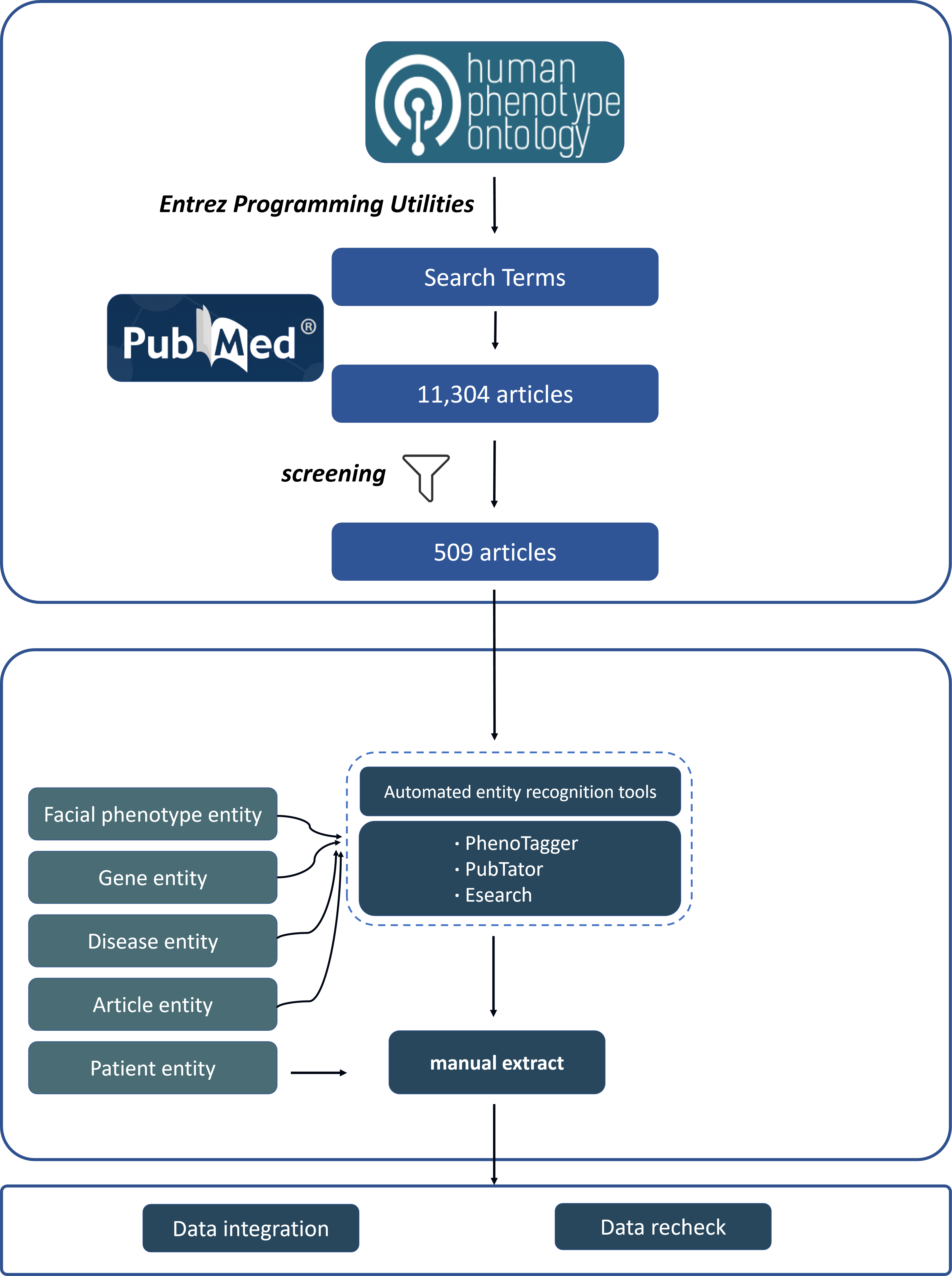}
    \caption{\textbf{The process of data collection. }}
    \label{fig process}
\end{figure}

\begin{table}[htbp]                    
    \caption{\textbf{Overview of data files and codes.}}        
    \label{tab overview}
    \centering      
    \resizebox{1.0\textwidth}{!}{                       
        \begin{tabular}{|c|c|c|}               
            \hline          
            \rowcolor{orange!50} 
            \textbf{File name}                           &\textbf{File type}                  &\textbf{File explanation} \\
            \hline
            FGDD\_labelled\_set                 &delimited text (.csv)      &The disease-labeled portion of the FGDD dataset includes 689 data records.  \\
            \hline
            FGDD\_unlabelled\_set               &delimited text (.csv)      &The unlabeled portion of the FGDD dataset includes 458 data records.  \\
            \hline
            phenotype                           &delimited text (.csv)      &Raw data, facial phenotype data collected in the publications.  \\
            \hline
            gene                                &delimited text (.csv)      &Raw data, gene data collected in the publications. \\
            \hline
            disease                             &delimited text (.csv)      &Raw data, disease data collected in the publications. \\
            \hline
            relation\_sample\_phenotype         &delimited text (.csv)      &Raw data, patients and their facial phenotypes.  \\
            \hline
            relation\_sample\_gene\_disease     &delimited text (.csv)      &Raw data, patients and their diseases, genetic variations. \\
            \hline   
            Baselines                           &IPython Notebook (.ipynb)  &Python codes, common classification algorithms trained on FGDD. \\        
            \hline   
            Explainability analysis             &IPython Notebook (.ipynb)  &Python codes, Explainability analysis from a global and local perspective. \\    
            \hline               
        \end{tabular}
    }
\end{table}

\begin{table}[htbp]                    
    \caption{\textbf{Performance of different classification algorithms on FGDD datasets.}}        
    \label{tab baseline}
    \centering      
    \resizebox{0.6\textwidth}{!}{                       
        \begin{tabular}{|c|c|c|}               
            \hline          
            \rowcolor{orange!50} 
            \textbf{Method}                &\textbf{Top-1 accuracy}   &\textbf{Macro-F1 score} \\                  
            Logistic Regression   &70.53	        &0.41 \\
            \hline
            Decision Tree	      &72.95	        &0.47\\
            \hline
            SVM	                  &64.25	        &0.34 \\
            \hline
            Random Forest	      &77.78	        &0.52 \\
            \hline
            Catboost	          &65.70	        &0.41 \\
            \hline
            Xgboost	              &64.73	        &0.35 \\
            \hline
            MLP	                  &\textbf{78.74}	  &\textbf{0.52} \\
            \hline               
        \end{tabular}
    }
\end{table}

\end{document}